\documentclass[12pt]{article}

\usepackage{epsfig}
\usepackage{amsbsy}
\usepackage{amssymb}
\usepackage{amsmath}

\begin{document}

\title{Potential and limitations of the HBT method}
\author{K.Zalewski
\\ Institute of Nuclear Physics Polish Academy of Sciences
\footnote{Address: Reymonta 4, 30 059 Krakow, Poland, e-mail:
zalewski@th.if.uj.edu.pl. This work has been partly supported by the Polish
Ministry of Science and Higher Education grant N N202 125437(2009-2012). }}
\maketitle

\begin{abstract}
The \textit{HBT} method is used to get information about the sizes, shapes and
sometimes also about the time evolution of the homogeneity regions in
hadroproduction processes. Homogeneity region $\textbf{K}$ is the region, where
the hadrons with momentum $\textbf{K}$ are produced. The shape and size of
homogeneity region $\textbf{K}$ is described by the Wigner function
$W(\textbf{K},\textbf{X})$ evaluated in the interaction representation after
all the hadrons had been produced. Additional information about the evolution
in time is contained in the emission function $S(K,X)$. A theorem is presented
and discussed which specifies which of the parameters characterizing the Wigner
function can and which cannot be measured using the \textit{HBT} method. In
order to obtain the complete Wigner function additional assumptions are needed.
For instance, it is enough to know the distribution of the centers of the
homogeneity regions $\langle \textbf{X} \rangle_\textbf{K}$. In order to find
the emission function further assumptions are required.  No systematic analysis
is available, but some instructive examples are discussed.
\end{abstract}
\noindent PACS numbers 25.75.Gz, 13.65.+i \\Bose-Einstein correlations,
interaction region determination. \vspace{0.5in}

\section{Introduction}

The \textit{HBT} method is a somewhat controversial \cite{KOP75} name for the
use of Bose-Einstein correlations to get information about the interaction
regions, defined below, in multiparticle production processes. For a recent
review see \cite{LIS}. In spite of its long history since the seminal paper of
the Goldhabers Lee and Pais \cite{GGLP}, the method remains a field of active
research. In the present paper we discuss what can and what cannot be measured
using the \textit{HBT} method. We also describe some open problems.

The interaction region is interpreted as follows. Consider a high energy
collision of two heavy nuclei where many hadrons are produced. The trajectory
of each hadron begins somewhere - in the point where the hadron got created.
This is a classical picture. It can be improved by replacing the points of
creation by small regions in space. All these points, or regions, averaged over
many similar collisions, form the interaction region. Thus, the interaction
region is, in general, not the volume where the fluid (strongly interacting
quark-gluon plasma?) created in the collision evolves. For instance, in some
models it is a two-dimensional shell.

If we were not constrained by the laws of Nature, the best description of the
interaction region would be the time dependent probability distribution in
phase space, i.e. the probability density for a particle of momentum
$\textbf{p}$ to be created at space-time point $X$. Since, however, according
to quantum mechanics it is not possible to measure simultaneously and precisely
the position and the momentum of a particle, one must compromise. In ordinary
quantum mechanics, where the number of particles is conserved, the Wigner
function is considered to be the best replacement for the phase space
probability distribution. In hadroproduction the situation is more complicated,
because the number of hadrons increases from zero to some final number. The
analogue of the Wigner function applicable in this case is known as the
emission function. It is implicit already in the work of Shuryak \cite{SHU},
but has been first explicitly defined and used by Pratt \cite{PRA}. For a
fairly recent discussion see \cite{LIS}.

\section{Emission function}

The emission function can be defined\footnote{In order to make the emission
function closer to the Wigner function, we have changed the sign of the
exponent with respect to the one given in \cite{LIS}} by the formula (cf. e.g.
\cite{LIS})

\begin{equation}\label{emi}
  S(p,X) = \int\!\!d^4Y\;\sum_RT_R^*(X + \frac{Y}{2})T_R(X -
  \frac{Y}{2})e^{-ipY},
\end{equation}
where

\begin{equation}\label{}
  X = \frac{1}{2}(x_1 + x_2),\quad Y = x_1 - x_2.
\end{equation}
$T_R(Z)$ is the probability amplitude for producing a hadron at space-time
point $Z$. This, of course, depends on the state of the surrounding, denoted
here by $R$, of $Z$ . By analogy with thermodynamics, the averaging over $R$ is
done on the product $T_R^*T_R$ and not on the single probability amplitudes
$T_R$. The integrand depends on two moments of time: $X^0 \pm \frac{1}{2}Y^0$.
Sometimes (see e.g. \cite{PRA}) incoherence in time of the production process
is assumed:

\begin{equation}\label{nocoht}
\sum_RT_R^*(X + \frac{Y}{2})T_R(X -
  \frac{Y}{2}) =  \delta(Y^0)\overline{\Sigma}(X,\textbf{Y}).
\end{equation}
Then there is only one time, occurring on both sides of equation (\ref{emi}),
and the emission function becomes closer to the Wigner function (see Section
4).

Note that in (\ref{emi}) there is neither integration nor differentiation with
respect to the components of $p$. Thus, momentum appears only as a parameter.
This is related to the fact, discussed in the following section, that it is not
possible to measure the whole interaction region. The most one can hope for is
to measure the \textit{homogeneity regions}. Homogeneity region $\textbf{K}$ is
the region where the hadrons with momentum $\textbf{K}$ got created. Thus, the
problem is to find the profiles of the homogeneity regions\footnote{Here and in
the following $\sim$ means: equal up to a known proportionality constant
irrelevant for our discussion.}:

\begin{equation}\label{prohom}
  p_\textbf{K}(\textbf{x}) \sim \int_{-\infty}^{+\infty}\!\!dX^0\; S(K,X).
\end{equation}
This limitation of the $HBT$ method was noticed by Bowler \cite{BOW}. The name
homogeneity region was introduced by Sinyukow \cite{SIN}.

The emission function is related to the single particle density matrix in the
momentum representation by the formula

\begin{equation}\label{rho}
\rho(\textbf{K},\textbf{q}) \sim \int\!\!d^4X\;S(K,X)e^{iqX},
\end{equation}
where

\begin{equation}\label{}
  K = \frac{1}{2}(p_1 + p_2),\quad q = p_1 - p_2.
\end{equation}
The density matrix does not depend on time, because it refers to a time when
all the hadrons are already present and propagate freely (To some extent one
can include final state interactions see e.g. \cite{LIS}). It is written in the
interaction picture.

Formula (\ref{rho}) rises two problems. How to measure
$\rho(\textbf{K},\textbf{q})$? The diagonal elements are given by the single
particle momentum distribution, but the textbook advice to look for
distributions of other measurable quantities in order to find the out of
diagonal elements is inapplicable here, because momenta are all we know how to
measure. A brilliant partial solution to this problem \cite{GGLP} is discussed
in the following section. Since Wigner's function is the Fourier transform of
the density matrix, knowing the density matrix is enough to find the size and
shape of a homogeneity region. The emission function, however, contains
additional interesting information about the time evolution of the
hadronization process.

The second problem is, how to solve for $S$ equation (\ref{rho}) for a given
density matrix? At first sight it might seem that it is enough to invert the
Fourier transformation, but in order to do that one would have to know
$\rho(\textbf{K},\textbf{q})$ for all the four-vectors $q$ and not only for all
the three-vectors $\textbf{q}$ as is the case. Here not much is known. We
discuss the problem in Section 4.

An important assumption concerning the emission function is the
\textit{smoothness assumption}, see e.g. \cite{LIS}. According to this
assumption, the dependence of $S(K,X)$ on $K$ is so weak that we can
replace\footnote{Care is taken to minimize the error. E.g. the product
$S(K,x_1)S(K,x_2)$ is replaced by $S(p_1,x_1)S(p_2,x_2)$, so that the errors of
the two substitutions partly cancel see \cite{LIS}.} $K$ by $p_1$ or $p_2$, or
replace $K_0$, by $\sqrt{m^2 + \textbf{K}^2}$, without changing significantly
the results. With this assumption many objections can be explained away. For
instance, two apparently very different versions of the $HBT$ method can be
shown to be equivalent \cite{LIS}, or the question can be answered: why the
arguments of the emission function, which are half sums just like the arguments
of the Wigner function, can be interpreted as particle momentum and position?
The smoothness assumption should hold for pairs of momenta $p_1,p_2$ important
for the analysis. For high energy collisions of heavy ions the relevant
momentum differences are small and the assumptions seems justified; for
$e^+e^-$ annihilations or $pp$ scattering they are much bigger and the
assumption is doubtful \cite{PRA2}. One of the outstanding open problems is,
why the $HBT$ method is applied with comparable success to heavy ion, $e^+e^-$
and $pp$ collisions?

\section{Measuring the density matrix}

Much information about the single particle density matrices can be obtained,
under certain assumptions, by studying the distributions of momenta for sets of
$n=1,2,\ldots $ identical mesons, for instance $\pi^-$ mesons \cite{GGLP}. The
formulae used to relate the $n$-particle momentum distributions
$P(\textbf{p}_1,\ldots,\textbf{p}_n)$ to the single particle density matrix
elements are:

\begin{eqnarray}\label{pnarho}
  P_1(\textbf{p}_1) =& N_1 \rho(\textbf{p}_1;\textbf{p}_1),\nonumber\\
  P_2(\textbf{p}_1,\textbf{p}_2) =& P(\textbf{p}_1)P(\textbf{p}_2) + N_{12}|\rho(\textbf{p}_1;\textbf{p}_2)|^2,\nonumber\\
  P_3(\textbf{p}_1,\textbf{p}_2,\textbf{p}_3) =& P(\textbf{p}_1)P(\textbf{p}_2)P(\textbf{p}_3) +
  P(\textbf{p}_1,\textbf{p}_2)P(\textbf{p}_3) +\\
P(\textbf{p}_2,\textbf{p}_3)P(\textbf{p}_1) +&
P(\textbf{p}_3,\textbf{p}_1)P(\textbf{p}_2) +
N_{123}\Re[\rho(\textbf{p}_1;\textbf{p}_2)\rho(\textbf{p}_2;\textbf{p}_3)\rho(\textbf{p}_3;\textbf{p}_1)]\;\;\;\nonumber
\end{eqnarray}
and so on, where $\Re$ stand for \textit{real part of} and $N_\alpha$ are
normalization constants, irrelevant for our discussion. The first formula
follows from the definition of the density matrix. The others are derived like
the second one, which was obtained, for a specific model and in a different
notation, in \cite{GGLP}: For two totaly uncorrelated particles, the
probability distribution for their momenta would be proportional to
$\rho(\textbf{p}_1;\textbf{p}_1) \rho(\textbf{p}_2;\textbf{p}_2)$; when the two
particles are identical and have spin zero, symmetrization introduces the
correction term $\rho(\textbf{p}_1;\textbf{p}_2)\rho(\textbf{p}_2;\textbf{p}_1)
= |\rho(\textbf{p}_1;\textbf{p}_2|^2$.

There is a problem with the  consistency of equations (\ref{pnarho}). Suppose
that exactly two particles are produced. Then $P_2(\textbf{p}_1,\textbf{p}_2)$
integrated over $\textbf{\textbf{p}}_2$ should give $P_1(\textbf{p}_1)$.
Actually, the integral of the first term on the right-hand-side of the formula
for $P_2$ yields $P_1$ and the second term supplies an unwanted correction. The
dependence of this correction on $\textbf{p}_1$ is usually different than that
in $P_1(\textbf{p}_1)$. Thus, it is not possible to compensate it by a
normalizing factor. The only way out is to make this correction negligibly
small. Since $\rho(\textbf{p}_1;\textbf{p}_1) $is known and fixed, one must
assume that $\rho(\textbf{p}_1;\textbf{p}_2)$ decreases rapidly with increasing
$|\textbf{q}|$.

The model described here has a number of other difficulties which, however, can
be relieved by a suitable handling of the experimental data and/or of the
comparison between theory and experiment. We will describe them very briefly. A
much more detailed description with many references can be found in the review
\cite{LIS}. The model assumes no correlations except the Bose-Einstein
correlations. The remedy is to construct a sample which contains all the
correlations except the Bose-Einstein correlations and compare it with the full
sample which contains all the correlations including the Bose-Einstein
correlations. The relative momentum $\textbf{q}$ of two charged particles with
similar momenta is strongly affected by Coulomb, and in some situations also by
strong, interactions. This is the famous problem of final state interactions
and is usually handled by introducing a suitable $\textbf{q}$-dependent
correction term. Many final hadrons are secondaries originating from decays of
resonances. The short-lived resonances are no problem, but the long-lived ones
mimic much bigger interaction regions. Actually, they produce in the ratio
$\frac{P_2(\textbf{p}_1,\textbf{p}_2)}{P_1(\textbf{p}_1)P_2(\textbf{p}_2)}$ a
narrow peak at small $|\textbf{q}|$ which is below experimental resolution and
is usually corrected for by changing the normalization factor $N_{12}$.
Finally, there are purely experimental problems like correcting for momentum
resolution, particle misidentification etc. We assume in the following that the
consistency condition is satisfied and that the experimental data have been
fully corrected, so that relations (\ref{pnarho}) can be applied.

The obvious question is, can one solve equations (\ref{pnarho}) for the single
particle density matrix $\rho(\textbf{p}_1;\textbf{p}_2)$? The answer is
negative. A simple calculation \cite{BIZ}, \cite{ZAL} shows  that the predicted
momentum distributions (for n=1,2,\ldots) do not change when
$\rho(\textbf{p}_1;\textbf{p}_2)$ changes as follows:

\begin{eqnarray}\label{ccon}
  \rho(\textbf{p}_1;\textbf{p}_2)&  \rightarrow&
  \rho(\textbf{p}_2;\textbf{p}_1)\quad\mbox{and/or}\\
  \label{phafac}\rho(\textbf{p}_1;\textbf{p}_2)&  \rightarrow&
 e^{if(\textbf{p}_1)} \rho(\textbf{p}_1;\textbf{p}_2)e^{-if(\textbf{p}_2)},
\end{eqnarray}
where $f(\textbf{p})$ is an arbitrary real function of a single momentum. This
group of transformations includes all the modifications of
$\rho(\textbf{p}_1;\textbf{p}_2)$ which affect none of the momentum
distributions. The first ambiguity is not much of a problem. It corresponds to
the space inversion of the homogeneity region. If the homogeneity region is
symmetric this has no effect. If it is not symmetric, one usually can find a
physical argument to chose one of the two possibilities. The second ambiguity,
however, is much more serious. Let us take an example. Choosing

\begin{equation}\label{exampl}
f(\textbf{p}) = \frac{1}{2}\sum_{i=x,y,z} a_ip_i^2,
\end{equation}
where $a_i$ are arbitrary real numbers, one finds that this implies the
modification

\begin{equation}\label{}
  S(K,X) \rightarrow S(K,\textbf{X} + \textbf{a}(\textbf{K}),X^0),
\end{equation}
where the components of $\textbf{a}(\textbf{K})$ are $a_iK_i$. Thus, every
homogeneity region, except the $\textbf{K} = \textbf{0}$ one, gets shifted.
Since the possible shifts are a class of functions of $\textbf{K}$, the
relative positions of the homogeneity regions can be almost arbitrarily
changed. This, incidentally, is a proof that at best one can hope to measure
the sizes and shapes of the individual homogeneity regions. Putting all the
homogeneity regions on top of each other, one could get a lower bound for the
size of the overall interaction region, but there is no upper bound.

The previous example is a very special case of a general theorem stating what
can and what cannot be measured using the \textit{HBT} method \cite{ZAL2}. The
formulation uses cumulants, so let us quote their definition. For the profile
of any homogeneity region $p_\textbf{K}(\textbf{x})$ we can define its
characteristic function $\langle e^{i\textbf{t}\cdot \textbf{x}} \rangle$ which
is, of course, a function of the vector $\textbf{t}$. Consider now the power
series expansion

\begin{equation}\label{}
  \log\langle
e^{i\textbf{t}\cdot \textbf{x}} \rangle =
\sum_{n_x,n_y,n_z}\;K(n_x,n_y,n_z)\frac{(it_x)^{n_x}}{n_x!}\frac{(it_y)^{n_y}}{n_y!}\frac{(it_z)^{n_z}}{n_z!},
\end{equation}
where the summation is over three sets of all non-negative integers. The
coefficients $K(n_x,n_y,n_z)$ are the cumulants of the probability distribution
$p_\textbf{K}(\textbf{x})$. The number $n_x+n_y+n_z$ is the order of the
cumulant. The cumulant is even (odd) when its order is even (odd).

The theorem consists of three points:
\begin{itemize}
  \item Every even cumulant of $p_\textbf{K}(\textbf{x})$ can be measured.
  \item No odd cumulant of $p_\textbf{K}(\textbf{x})$ can be measured.
  \item When the first order cumulants, i.e. $\langle \textbf{X}
  \rangle_\textbf{K}$, are given, they fix the phase $f(\textbf{p})$ up to an irrelevant
  constant and, for $\langle \textbf{X} \rangle_\textbf{K} \neq \textbf{0}$,  eliminate the freedom of inversion.  Thus, the full $p_\textbf{K}(\textbf{x})$
  become measurable.
\end{itemize}

Let us consider a few applications of this theorem. The second order cumulants
are elements of the covariance matrix $\langle(x_i - \langle x_i \rangle)(x_j -
\langle x_j \rangle) \rangle_\textbf{K}$ where $x_i = x,y,z$. Experimentalist
routinely measure the \textit{HBT} radii and use them to calculate the elements
of the covariance matrices of the homogeneity regions. The theorem shows that
this interpretation is not affected by the ambiguity (\ref{phafac}). In the
imaging method (see \cite{DAP} and references given there) one determines from
the measured momentum distributions the probability distributions
$\overline{p}_\textbf{K}(\textbf{x}_1 - \textbf{x}_2)$ for pairs of points
within a homogeneity region. This is much less information than given by the
profiles of the homogeneity regions. For instance, the distributions
$\overline{p}_\textbf{K}$ are not sensitive to the averages $\langle \textbf{X}
\rangle_\textbf{K}$. However, one finds that the distribution
$\overline{p}_\textbf{K}$ depends only on the even cumulants of the
distribution $p_\textbf{K}$ and that it depends on all of them. Therefore, it
is reliably measurable and it gives much more information than the measurements
of the \textit{HBT} radii.

On the other hand, the centers of the homogeneity regions $\langle
\textbf{X}\rangle_K$ are first order cumulants and cannot be reliably measured.
This is a much shorter derivation of the conclusion from example (\ref{exampl})
that the relative positions of the homogeneity regions are not measurable.

Of special interest is the third point. The collisions of heavy ions are
usually described using rather simple-minded tools, like Euler's hydrodynamics.
It is much easier to believe that such analyses give reasonably the centers of
the homogeneity regions than that they reproduce all the intricacies of their
shapes. According to the third point, however, once the distribution of the
centers is know, everything else can be unambiguously measured without further
model assumptions.

\section{Emission function and Wigner function}

Much less work has been done on the ambiguities in the solution of equation
(\ref{rho}) for $\rho(\textbf{K},\textbf{q})$ known. Instead, one usually
derives from a model, or guesses, an emission function and checks whether it
yields the momentum distributions in agreement with experiment. Since a general
analysis has not yet been performed, we present here only some partial results.

Let us assume that in some reference frame all the hadrons got created
simultaneously at time $t=0$. Then the emission function can be written in the
form

\begin{equation}\label{}
  S(K,X) = \delta(t)\overline{S}(K,X),
\end{equation}
where the function $\overline{S}(K,X)$ should be found from equation
(\ref{rho}). Substituting into (\ref{rho}), one finds, a unique solution

\begin{equation}\label{deltat}
  S(K,X) \sim \delta(t)W(\textbf{K},\textbf{X}),
\end{equation}
where $W$ denotes the Wigner function. Thus, in this case the emission function
is as good as the Wigner function. A solution of this type always exists,
assuming that the hadronization process is over at $t=0$, but usually it is not
the realistic solution we are looking for.

Assuming (in some reference frame) no coherence in time, one can relate the
emission function to the Wigner function by the formula \cite{ZAL3}

\begin{equation}\label{wdift}
S(K,X) \sim \frac{\overline{W}_{dt}(\textbf{K},X)}{dt}.
\end{equation}
$\overline{W}_{dt}$ is the Wigner function for the particles produced in the
time interval $dt$ around $t = X^0$. This Wigner function is normalized to the
number of particles produced in the time interval $dt$ and not to one as is
usual for Wigner functions. To mark this difference we introduced the overline.
With this normalization $\overline{W}_{dt}$ tends to zero with $dt \rightarrow
0$ and it has to be divided by $dt$ to give a finite result. Note that this
solution for $S(K,X)$ does not depend on $K^0$.  The equation for $S(K,X)$ can
by obtained by integrating both sides of (\ref{wdift}) over time. The result is

\begin{equation}\label{wnaints}
  W(\textbf{K},X)  \sim \int\!\!dt\;S(K,X).
\end{equation}
Solution (\ref{wdift}) is one among the infinity of solutions of
(\ref{wnaints}) and the \textit{HBT} method gives no hint how to find it.

However, (\ref{wnaints}) can be used as a sum rule for $S(K,X)$.  If
assumptions are made about the time dependence of $S(K,X)$, it is possible to
get one parameter of the assumed time distribution from experiment. An example
is contained in \cite{KOP73}. This paper proposed a once very famous formula to
measure both the size and the life-time of the interaction region by the
\textit{HBT} method. The formula was used for years by many experimental
groups. The authors assumed that all the particle sources are produced
simultaneously at $t=0$, that there is no coherence in time and that each
source decays exponentially with the same life time. Then, in agreement with
our analysis, they were able to determine this life time of the sources.

Let us consider a generalization of the previous case. Suppose that $S(K,X)$
does not depend on $\textbf{q}^2$. This is more general than the previous
assumption that is does not depend on $K^0$. For instance, it is enough to
assume that, in the spirit of the smoothness assumption, $K^0$ in $S(K,X)$ can
be replaced by  $\sqrt{m^2 + \textbf{K}^2}$. Applying the Fourier
transformation to both sides of equation (\ref{rho}) we convert the density
matrix into the Wigner function and using the identity

\begin{equation}\label{}
Kq = 0 = K^0q^0 - \textbf{K}\cdot \textbf{q},
\end{equation}
to eliminate $q^0$, we find after trivial integrations

\begin{equation}\label{}
  W(\textbf{K},\textbf{X}) \sim \int\!\!dt\;S(K,\textbf{X} + \frac{\textbf{K}}{K^0}t,t)
\end{equation}
In the $\textbf{K} = \textbf{0}$ frame one recovers formula (\ref{wnaints}).
Actually, the use of this frame has a number of advantages (see e.g.
\cite{LIS}) and should be strongly recommended.

\section{Conclusions}

It is not possible to measure the size and/or shape of the full interaction
region using the \textit{HBT} method. Therefore, efforts concentrate on
measurements of the homogeneity regions. From the classical point of view, the
best description of the size and shape would be a probability distribution for
producing a particle with momentum $\textbf{K}$ at space-time point $X$. As
known from quantum mechanics, however, this is not possible. The usual
translation into quantum physics is to look for the Wigner function
$W(\textbf{K},\textbf{X})$. When written in the interaction picture and after
all the hadrons have been produced, this Wigner function does not depend on
time. Its interpretation as a probability distribution
$p_\textbf{K}(\textbf{X}) = W(\textbf{K},\textbf{X})$ is only an approximation,
e.g. because Wigner functions can be negative in certain regions. It seems,
however, that at least for high-energy heavy ion scattering this approximation
is reasonable (for a recent discussion see \cite{ZAL4}).

The theorem presented and discussed in Section 3 states that: The even
cumulants of $p_\textbf{K}(\textbf{X})$ can be measured using the \textit{HBT}
method. This, in particular, justifies the measurements of the \textit{HBT}
radii and the imaging method. The odd cumulants cannot be measured, which
explains, in particular, why one has to study the separate homogeneity regions
instead of the whole interaction region. The theorem also shows that when the
distribution $\langle \textbf{X} \rangle_\textbf{K}$ is known, it is possible
to measure the full Wigner functions for all the homogeneity regions and
consequently also the full interaction region.

In order to learn about the time evolution of the homogeneity regions one uses
emission functions. Even when the Wigner function is known, the evaluation of
the emission function requires additional assumptions. When these assumptions
are too strong, one obtains no more than the input. This is illustrated by the
example leading to formula (\ref{deltat}). With a judicious choice of
assumptions, however, one can obtain additional information. An example is the
model of Kopylov and Podgoretskii \cite{KOP73},  where the life time of the
hadron sources can be measured. Of course, such results are no more credible
than the assumptions used to derive them.

Nowadays people usually propose full models of hadronization, which among other
things predict also the results of the \textit{HBT} measurements. The
comparison with these experimental measurements is used to test the model and
sometimes also to fix some of its parameters. The credibility of the model, and
consequently of the description of the interaction region that it offers, is
based on a comparison of its prediction with a variety of experimental data and
not just on the study of the \textit{HBT} results. The question: how much one
can learn about the interaction region in a (as nearly as possible)
model-independent way? remains, however, interesting.

\end{document}